\documentclass[a4paper,12pt]{article}  
\usepackage{color}
\usepackage{epsfig}
\usepackage{amsfonts}
\usepackage{amsmath}
\usepackage{amssymb, amsbsy}
\usepackage{mathtools}
\usepackage{setspace}
\usepackage{chngpage}
\usepackage{caption}
\usepackage{graphicx}
\usepackage[singlelinecheck=false]{caption} 

\usepackage[para,online,flushleft]{threeparttable} 
\makeatletter
\def\TPT@doparanotes{\par
   \prevdepth\z@ \TPT@hsize
   \TPTnoteSettings
   \parindent\z@ \pretolerance 8
   \linepenalty 200
   \renewcommand\item[1][]{\relax\ifhmode \begingroup
       \unskip
       \advance\hsize 10em 
       \penalty -45 \hskip\z@\@plus\hsize \penalty-19
       \hskip .15\hsize \penalty 9999 \hskip-.15\hsize
       \hskip .01\hsize\@plus-\hsize\@minus.01\hsize 
       \hskip 6em\@plus .3em
      \endgroup\fi
      \tnote{##1}\,\ignorespaces}%
   \let\TPToverlap\relax
   \def\endtablenotes{\par}%
}

\usepackage[english]{babel} 
\usepackage{ragged2e}
\usepackage{blindtext}

\usepackage[hidelinks]{hyperref}
\usepackage{xcolor}
\hypersetup{
    colorlinks,
    linkcolor={red!50!black},
    citecolor={blue!50!black},
    urlcolor={blue!80!black}
}

\usepackage[square, authoryear]{natbib}
\usepackage{url}
\DeclareMathOperator{\se}{se}

\DeclareMathOperator*{\cor}{cor}

\makeatletter
\renewcommand\@biblabel[1]{}
\makeatother

\begin{document}
\title{Disentangling the effects of traits with shared clustered genetic predictors using multivariable Mendelian randomization}
\author{Fatima Batool\textsuperscript{1}\thanks{Email: fatima.batool@mrc-bsu.cam.ac.uk, MRC Biostatistics Unit, East Forvie Site, Robinson Way, Cambridge, CB2 0SR, UK. Telephone: +44 1223 (7)60716.} \and Ashish Patel\textsuperscript{1} \and Dipender Gill\textsuperscript{2,3,4} \and Stephen Burgess\textsuperscript{1,5}\thanks{Email: sb452@medschl.cam.ac.uk, MRC Biostatistics Unit, Forvie Site, Robinson Way, Cambridge, CB2 0SR, UK. Telephone: +44 1223 748651. } \\ \\
\textsuperscript{1}  MRC Biostatistics Unit,  Institute of Public Health, Biomedical Campus, \\ University of Cambridge, Cambridge, UK \\~~\\
\textsuperscript{2} Department of Epidemiology and Biostatistics, School of Public Health, \\ Imperial College London, London, UK\\ ~~\\
\textsuperscript{3} Genetics Department, Novo Nordisk Research Centre,  Oxford, UK \\ ~~\\
\textsuperscript{4} Clinical Pharmacology and Therapeutics Section, Institute for Infection and \\ Immunity, St George’s, University of London, London, UK.\\~~\\
\textsuperscript{5} Cardiovascular Epidemiology Unit, Department of Public Health and \\ Primary Care,  University of Cambridge, Cambridge, UK }
\maketitle

\noindent This research was funded by United Kingdom Research and Innovation Medical Research Council (MC$\_$UU$\_$00002$\_$7 and MR/S037675/1) and supported by the National Institute for Health Research Cambridge Biomedical Research Centre (BRC-1215-20014). Fatima Batool and Stephen Burgess are supported by a Sir Henry Dale Fellowship jointly funded by the Wellcome Trust and the Royal Society (204623$/$Z$/$16$/$Z). The views expressed are those of the authors and not necessarily those of the National Institute for Health Research or the Department of Health and Social Care. Dr Gill is supported by the British Heart Foundation Centre of Research Excellence (RE/18/4/34215) at Imperial College London and a National Institute for Health Research Clinical Lectureship at St. George's, University of London (CL-2020-16-001).

\vspace{3mm}

\noindent The authors have expressed no conflict of interest besides Dr Gill is employed part-time by Novo Nordisk. 
\subsection*{Abstract}
When genetic variants in a gene cluster are associated with a disease outcome, the causal pathway from the variants to the outcome can be difficult to disentangle. For example, the chemokine receptor gene cluster contains genetic variants associated with various cytokines. Associations between variants in this cluster and stroke risk may be driven by any of these cytokines. Multivariable Mendelian randomization is an extension of standard univariable Mendelian randomization to estimate the direct effects of related exposures with shared genetic predictors. However, when genetic variants are clustered, a Goldilocks dilemma arises: including too many highly-correlated variants in the analysis can lead to ill-conditioning, but pruning variants too aggressively can lead to imprecise estimates or even lack of identification. We propose multivariable methods that use principal component analysis to reduce many correlated genetic variants into a smaller number of orthogonal components that are used as instrumental variables. We show in simulations that these methods result in more precise estimates that are less sensitive to numerical instability due to both strong correlations and small changes in the input data. We apply the methods to demonstrate the most likely causal risk factor for stroke at the chemokine gene cluster is monocyte chemoattractant protein-1.

\vspace{3mm}

\noindent \noindent \textbf{Key words:} Mendelian randomization, gene cluster, correlated variants, dimension reduction, causal inference.

\clearpage

\section*{Introduction}
Genetic variants associated with molecular and phenotypic traits can provide evidence on the causal pathways linking the associated trait with a disease outcome \citep{burgess2017gwas}. Various analytical approaches, including Mendelian randomization and colocalization, have been proposed that synthesize data on genetic associations to assess the nature of the relationship between a trait and a disease. In Mendelian randomization, it is assumed that the only pathway by which selected genetic variants influence the outcome is via the associated trait \citep{lawlor2007}. Formally, genetic variants are assumed to satisfy the assumptions of an instrumental variable \citep{didelez2007}. If multiple independent variants associated with the same trait show a consistent pattern of associations with the outcome, then it is plausible that the trait has a causal effect on the outcome \citep{bowden2015median, lawlor2016triangulation}.

We here consider an extension to standard Mendelian randomization known as multivariable Mendelian randomization, which allows genetic variants to be associated with multiple related traits \citep{burgess2014pleioaje}. For instance, it is difficult to find specific genetic predictors of fat-free mass that are not also associated with fat mass. Multivariable Mendelian randomization can be implemented by fitting a multivariable regression model using genetic associations with each of the traits as predictors \citep{sanderson2018}. Coefficients from this model represent direct effects; that is, the effect of varying one of the traits in the model while keeping other traits constant \citep{burgess2017summnetwork, carter2019}. Such investigations have suggested that fat mass rather than fat-free mass influences cardiovascular disease risk \citep{larsson2020body}, and that, amongst a set of lipid traits, apolipoprotein B is the primary driver of coronary heart disease risk \citep{zuber2021}.

Often, Mendelian randomization investigations are conducted using genetic variants from a single genetic region, an approach known as cis-Mendelian randomization \citep{schmidt2020}. This approach is particularly common when the risk factor is a gene product, such as gene expression or circulating levels of a protein. Such analyses are somewhat fragile, as the evidence is based on a single genetic region and so it is not possible to assess heterogeneity of findings across multiple genetic regions that represent independent datapoints \citep{burgess2020guidelines}. However, if the function of the gene is well-understood, these analyses can be particularly insightful into the impact of intervening on a specific biological pathway. In some cases, the function of the gene may correspond to the action of a pharmacological agent, and hence the analysis is informative about a druggable pathway \citep{gill2021wellcome}. Examples include the use of variants in the \emph{HMGCR} gene region to inform about the impact of statin drugs \citep{ference2015}, and variants in the \emph{IL6R} gene region about the impact of interleukin-6 receptor inhibitors, such as tocilizumab \citep{sarwar2012}.

However, some genetic regions contain multiple genes (referred to as a gene cluster), and so are associated with multiple gene products. For example, the \emph{FADS} locus contains genetic predictors of various fatty acids \citep{lattka2010}, and the \emph{IL1RL1}–-\emph{IL18R1} locus (the interleukin-1 receptor cluster) contains protein quantitative trait loci (pQTLs) for several proteins \citep{sun2017}. While variants in the interleukin-1 cluster are associated with several autoimmune diseases \citep{timms2004, zhu2008}, it is difficult to determine which of the proteins are causal risk factors \citep{reijmerink2010}. Although a multivariable cis-Mendelian randomization approach has been proposed to disentangle complex gene regions and identify the causal drivers of disease \citep{porcu2019}, authors of this approach suggest pruning genetic variants to near independence ($r^2 < 0.1$) to avoid potential problems of collinearity. However, it may not be possible to find sufficient near-independent variants for the multivariable regression model to give precise estimates for each trait. While it is possible to prune at a less strict threshold, we have previously shown that under-pruning can result in ill-conditioning \citep{burgess2017fine}. This represents a Goldilocks dilemma: too much pruning and we get imprecision or even lack of identification; too little pruning and we can get results that are highly sensitive to small changes in the estimated correlation matrix, and can be nonsensical.

We propose two methods for multivariable cis-Mendelian randomization that perform dimension reduction on the available genetic variants at a locus using principal component analysis (PCA). These methods reduce information on large numbers of highly correlated variants into a smaller number of orthogonal components, allowing efficient multivariable analyses to be implemented that are not so sensitive to high correlations between variants or small changes in the inputs. We demonstrate the superiority of these methods over pruning methods in a simulation study, and illustrate the methods in a applied analysis investigating effects on stroke risk of three cytokines associated with a gene cluster on chromosome 17.

\section*{Methods}
\subsection*{Overview of the approach}
Multivariable Mendelian randomization takes genetic variants that are each associated with at least one of a set of related exposure traits, and satisfy the instrumental variable assumptions for multivariable Mendelian randomization:
\begin{itemize}
\item[i.] each variant is associated with one or more of the exposures,
\item[ii.] each exposure is associated with one or more of the genetic variants,
\item[iii.] each variant is not associated with the outcome via a confounding pathway, and
\item[iv.] each variant does not affect the outcome directly, only possibly indirectly via one or more of the exposures.
\end{itemize}
Although the approach was originally developed for use with individual-level data using the established two-stage least squares method \citep{burgess2014pleioaje}, equivalent estimates can be obtained using summarized genetic association data that are typically reported by genome-wide association studies (GWAS) \citep{sanderson2018}. We use summarized genetic association data \citep{bowden2017}, and denote the genetic association of variant $j$ with exposure trait $k$ as $\hat{\beta}_{Xjk}$; this is the beta-coefficient from univariable regression of the trait on the variant. We denote the genetic association of variant $j$ with the outcome as $\hat{\beta}_{Yj}$ and its standard error as $\se(\hat{\beta}_{Yj})$; again, this is obtained from regression of the outcome on the variant.

\subsection*{Inverse-variance weighted method}
If the genetic variants are uncorrelated, then multivariable Mendelian randomization estimates can be obtained by fitting a multivariable model using weighted linear regression:
\begin{equation}
\hat{\beta}_{Yj} = \theta_{1} \; \hat{\beta}_{Xj1} + \theta_{2} \; \hat{\beta}_{Xj2} + \ldots + \theta_{K} \; \hat{\beta}_{XjK} + \epsilon_{j} \qquad \epsilon_{j} \sim \mathcal{N}(0, \se(\hat{\beta}_{Yj})^2)
\end{equation}
for variants $j = 1, 2, \ldots, J$, where $K$ is the total number of traits ($K > J$), and the error terms $\epsilon_{j}$ have independent normal distributions \citep{burgess2014pleioajeappendix}. The parameter $\theta_k$ is an estimate of the direct effect of the $k$th trait on the outcome (that is, the effect of intervening on that trait while keeping all other traits unchanged) \citep{carter2019}. We refer to this method as the multivariable inverse-variance weighted (MV-IVW) method, as it is an extension of the univariable IVW method \citep{burgess2013genepi} to the multivariable setting.

If the genetic variants are correlated, then we allow the error terms to be correlated and use generalized weighted linear regression:
\begin{equation}
\mathbf{\hat{\boldsymbol\beta}_{Y}} = \theta_{1} \; \mathbf{\hat{\boldsymbol\beta}_{X1}} + \theta_{2} \; \mathbf{\hat{\boldsymbol\beta}_{X2}} + \ldots + \theta_{K} \; \mathbf{\hat{\boldsymbol\beta}_{XK}} + \mathbf{\boldsymbol\epsilon} \qquad \mathbf{\boldsymbol\epsilon} \sim \mathcal{N}(\boldsymbol0, \Sigma)
\end{equation}
where bold face represents vectors, and $\Sigma$ is a variance-covariance matrix with elements $\Sigma_{j_1, j_2} = \se(\hat{\beta}_{Yj_1}) \; \se(\hat{\beta}_{Yj_2}) \; \rho_{j_1, j_2}$, with $\rho_{j_1, j_2}$ representing the correlation between the $j_1$th and $j_2$th variants. This method was advocated by Porcu \emph{et al} \citep{porcu2019} for the analysis of summarized genetic association data on gene expression traits with shared genetic predictors. Estimates are obtained as:
\begin{equation}
\mathbf{\hat{\boldsymbol\theta}_{MV-IVW}} = (\mathbf{\hat{\boldsymbol\beta}_{X}}^T \Sigma^{-1} \mathbf{\hat{\boldsymbol\beta}_{X}})^{-1} \mathbf{\hat{\boldsymbol\beta}_{X}}^T \Sigma^{-1} \mathbf{\hat{\boldsymbol\beta}_{Y}}
\end{equation}
where $\mathbf{\hat{\boldsymbol\beta}_{X}}$ is the $J$ by $K$ matrix of genetic associations with the traits, and $\mathbf{\hat{\boldsymbol\beta}_{Y}}$ is the $J$ by $1$ vector of genetic associations with the outcome.

This calculation requires inversion of the variance-covariance matrix $\Sigma$, which can lead to numerical instability if the matrix of correlations between genetic variants is near-singular. This occurs when there is a set of genetic variants that is close to being linearly dependent. If a set of genetic variants is linearly dependent (that is, there is at least one variant that can be perfectly predicted based on the other variants), then the correlation matrix will be exactly singular, and so cannot be inverted. If a set of genetic variants is almost but not exactly linearly dependent, then the correlation matrix can be inverted, but some elements of the matrix inverse will be very large in magnitude. This results in an ill-conditioned problem, meaning that small changes in the inputs can lead to large changes in the estimates. The condition number of a matrix is a measure of ill-conditioning; for a positive-definite symmetric matrix, this can be calculated as the ratio of the largest to the smallest eigenvalue. While there are no universal thresholds, condition numbers over 100 are cause for concern, particularly if the genetic associations are known to a limited degree of precision.

To implement the proposed PCA method, we first consider the matrix $\Psi$ where:
\begin{equation}
\Psi_{j_1, j_2} = \sum_k |\hat{\beta}_{Xj_1k}| \; \sum_k |\hat{\beta}_{Xj_2k}| \; \se(\hat{\beta}_{Yj_1})^{-1} \; \se(\hat{\beta}_{Yj_2})^{-1} \; \rho_{j_1, j_2}.
\end{equation}
This is a weighted version of the variance-covariance matrix, with weights taken as the sum of the absolute values of the genetic associations with the traits. Obtaining principal components of this matrix ensures that the top principal components assign greater weights for variants having larger associations with the traits and more precise associations with the outcome. Considering the PCA decomposition $\Psi = W \Lambda W^T$, where $W$ is the matrix of eigenvectors (or loadings) and $\Lambda$ is the diagonal matrix with the eigenvalues $\lambda_1 > \ldots > \lambda_J$ on the diagonal, let $W_k$ be the matrix constructed of the first $k$ columns of $W$. Then we define:
\begin{align*}
\mathbf{\tilde{\boldsymbol\beta}_X} &= W_k^T \mathbf{\hat{\boldsymbol\beta}_{X}} \mbox{ as the matrix of transformed genetic associations with the exposure traits} \\
\mathbf{\tilde{\boldsymbol\beta}_Y} &= W_k^T \mathbf{\hat{\boldsymbol\beta}_{Y}} \mbox{ as the vector of transformed genetic associations with the outcome} \\
\tilde{\Sigma} &= W_k^T \Sigma W_k \mbox{ as the transformed variance-covariance matrix.}
\end{align*}
The multivariable inverse-variance weighted principal component analysis (MV-IVW-PCA) estimate is given by:
\begin{equation}
\mathbf{\hat{\boldsymbol\theta}_{MV-IVW-PCA}} = (\mathbf{\tilde{\boldsymbol\beta}_{X}}^T \tilde{\Sigma}^{-1} \mathbf{\tilde{\boldsymbol\beta}_{X}})^{-1} \mathbf{\tilde{\boldsymbol\beta}_{X}}^T \tilde{\Sigma}^{-1} \mathbf{\tilde{\boldsymbol\beta}_{Y}}
\end{equation}
This is an adaptation of the MV-IVW method using transformed genetic instruments that represent linear weighted scores comprised of the original genetic variants, where the weights of the scores are the eigenvectors from the PCA decomposition. As the principal components are orthogonal, the transformed variance-covariance matrix should not be near-singular. The standard errors of these estimates are:
\begin{equation}
\se(\hat{\theta}_{k, MV-IVW-PCA}) = \sqrt{(\mathbf{\tilde{\boldsymbol\beta}_{X}}^T \tilde{\Sigma}^{-1} \mathbf{\tilde{\boldsymbol\beta}_{X}})^{-1}_{kk}}.
\end{equation}
In our investigations, we set the number of principal components to explain 99\% of the variance in the matrix $\Psi$.

\subsection*{Limited information maximum likelihood method}
An alternative method for instrumental variable analysis is the limited information maximum likelihood (LIML) method \citep{baum2007}. The estimate from the LIML method minimizes the Anderson--Rubin statistic \citep{anderson1949}, which is a measure of heterogeneity of the estimates based on the different genetic variants. Both the IVW and LIML methods are part of a larger family of methods, known as the generalized method of moments (GMM) \citep{hansen1982}. We here derive a multivariable analogue of the LIML method that can be implemented using summarized genetic association data for correlated variants. We refer to this method as the multivariable limited information maximum likelihood (MV-LIML) method.

For the MV-LIML method, we require the additional knowledge of the $K \times K$ correlation matrix of exposures, which we denote $\Phi$. In the simulation study, we set this matrix to be the identity.

We let $\mathbf{g}(\boldsymbol\theta)=\mathbf{\hat{\boldsymbol\beta}_{Y}}-\mathbf{\hat{\boldsymbol\beta}_{X}} \; \boldsymbol\theta$, where
$\boldsymbol\theta=(\theta_{1},\ldots,\theta_{K})^{T}$. Under the assumption that all $J$ variants are valid instruments, setting $\mathbf{g}(\boldsymbol\theta)=\boldsymbol0$ provides a set of $J$ estimating equations for the $K$ unknown parameters $\theta_k$. When $J>K$, if the genetic variants are linearly independent, it is generally not possible to find an estimator $\mathbf{\hat{\boldsymbol\theta}}$ that can set $\mathbf{g}(\hat{\boldsymbol\theta})=\boldsymbol0$. Thus, LIML-based methods take $K$ linear combinations of the $J$ estimating equations, where the weights in the linear combination are chosen to minimize the variance of the resulting estimator $\mathbf{\hat{\boldsymbol\theta}}$.

In particular, the MV-LIML estimator is given by
\begin{equation}
\mathbf{\hat{\boldsymbol\theta}_{MV-LIML}}=\arg\min_{\boldsymbol\theta}\hat{Q}(\boldsymbol\theta)
\end{equation}
where $Q(\boldsymbol\theta)=\mathbf{g}(\boldsymbol\theta)^{T} \; \Omega(\boldsymbol\theta)^{-1} \; \mathbf{g}(\boldsymbol\theta)$,
and $\Omega(\boldsymbol\theta)$ is a $J\times J$ matrix with its $(j_{1},j_{2})$th element given by
\begin{equation}
\Omega_{j_{1},j_{2}}(\theta)=\big(\text{se}(\hat{\beta}_{Y_{j_{1}}})\text{se}(\hat{\beta}_{Y_{j_{2}}})\rho_{j_{1},j_{2}}\big)+\sum_{k=1}^{K}\sum_{l=1}^{K}\sqrt{\text{se}(\hat{\beta}_{X_{j_{1}k}})\text{se}(\hat{\beta}_{X_{j_{2}k}})}\sqrt{\text{se}(\hat{\beta}_{X_{j_{1}l}})\text{se}(\hat{\beta}_{X_{j_{2}l}})}\rho_{j_{1},j_{2}}\Phi_{k,l}\theta_{k}\theta_{l}.
\end{equation}
For exposure $k$, the MV-LIML estimator of $\theta_{k}$ is given by the $k$th element of $\mathbf{\hat{\boldsymbol\theta}_{MV-LIML}}$,
and its standard error is given by $\sqrt{\hat{V}_{k,k}}$, where $\hat{V}_{k,k}$ is the $k$th diagonal element of the $K \times K$
matrix $\hat{V}=(\mathbf{\hat{\boldsymbol\beta}_{X}}^{T} \; \Omega(\mathbf{\hat{\boldsymbol\theta}_{MV-LIML}})^{-1} \; \mathbf{\hat{\boldsymbol\beta}_{X}})^{-1}$.

Theoretical results suggest LIML provides robust estimation when using many weak instruments \citep{chao2005}. For univariable cis-Mendelian randomization analyses, simulation evidence has further highlighted the low bias properties of LIML-based estimators in finite samples \citep{patel2020}.

We consider a version of the LIML method using PCA to perform dimension reduction on the set of genetic variants by replacing:
\begin{align}
\mathbf{\tilde{g}(\boldsymbol\theta)} &= W_k^T \mathbf{g}(\boldsymbol\theta) \\
\tilde{\Omega}(\boldsymbol\theta)     &= W_k^T \Omega(\boldsymbol\theta) W_k \notag \\
\tilde{Q}(\boldsymbol\theta)          &= \mathbf{\tilde{g}}(\boldsymbol\theta)^{T} \; \tilde{\Omega}(\boldsymbol\theta)^{-1} \; \mathbf{\tilde{g}}(\boldsymbol\theta) \notag
\end{align}
and minimizing $\tilde{Q}(\boldsymbol\theta)$, with $W_k$ defined as previously. We refer to this method as MV-LIML-PCA. As per the MV-IVW-PCA method, we set the number of principal components to explain 99\% of the variance in the matrix $\Psi$.

\subsection*{Simulation study}
We perform a simulation study to assess whether the proposed PCA methods are able to detect which out of a set of traits with shared clustered genetic predictors influences an outcome, and whether the causal effects of the traits can be estimated reliably.

We consider a scenario with three traits and 100 correlated genetic variants. We generate 10\thinspace000 simulated datasets according to the following data-generating model for 20\thinspace000 individuals indexed by $i$:
\begin{align}
  A_{j_1, j_2} &\sim \mbox{Uniform}(-0.3, 1) \mbox{ for } j_1, j_2 = 1, \ldots, 100 \notag \\
  B            &= \cor(A \; A^T) \notag \\
  \mathbf{G_i} &\sim \mathcal{N}_J(\boldsymbol0, B) \notag \\
  \alpha_j     &\sim \mathcal{N}(0.08, 0.01^2) \mbox{ for } j = 1, \ldots, 15 \notag \\
  X_{i1}       &= \sum_{j=1}^5     \alpha_j G_{ij} + U_{i1} + \epsilon_{Xi1} \notag \\
  X_{i2}       &= \sum_{j=6}^{10}  \alpha_j G_{ij} + U_{i2} + \epsilon_{Xi2} \notag \\
  X_{i3}       &= \sum_{j=11}^{15} \alpha_j G_{ij} + U_{i3} + \epsilon_{Xi3} \notag \\
  Y_{i}        &= 0.4 X_{i1} - 0.6 X_{i3} + U_{i1} + U_{i2} + U_{i3} + \epsilon_{Yi} \notag \\
  U_{i1}, U_{i2}, U_{i3}, \epsilon_{Xi1}, \epsilon_{Xi2}, \epsilon_{Xi3}, \epsilon_{Yi} &\sim \mathcal{N}(0, 1) \mbox{ independently} \notag
\end{align}

The genetic variants $G_j$ are simulated from a multivariable normal distribution with mean zero and variance-covariance matrix $B$. The traits $X_1$, $X_2$, and $X_3$ are simulated such that 5 variants influence the first trait, the next 5 influence the second trait, the next 5 influence the third trait, and the remaining 85 do not influence any trait. However, due to the moderately large correlations between the genetic variants (which typically range from around $-0.1$ to $+0.6$ with an interquartile range from around $+0.2$ to $+0.4$), typically each of the 100 variants is associated with all three traits at a genome-wide level of significance. The outcome $Y$ is influenced by traits $X_1$ and $X_3$, with the true effect of $X_1$ set at $+0.4$ and the true effect of $X_3$ set at $-0.6$. The associations between the traits and the outcome are affected by confounders $U_1$, $U_2$, and $U_3$.

We estimate genetic associations with the exposures on the first 10\thinspace000 individuals, and genetic associations with the outcome on the subsequent 10\thinspace000 individuals. Correlations between the genetic variants are estimated on the first 10\thinspace000 individuals. This represents a two-sample scenario, where genetic associations with the exposures and outcome are obtained on non-overlapping samples \citep{pierce2013}. The mean instrument strength based on the 5 causal variants for each trait is $R^2 = 3.5\%$, corresponding to a mean univariable F statistic (on 5 and 19\thinspace994 degrees of freedom) for each trait of around 145, and a conditional F statistic (on 15 and 19\thinspace984 degrees of freedom) for each trait of around 22 \citep{sanderson2018}.

We compare four different methods: the MV-IVW and MV-LIML methods with various choices of genetic variants as inputs, and the MV-IVW-PCA and MV-LIML-PCA methods described above. For the MV-IVW and MV-LIML methods, we consider pruning the variants at thresholds of $|\rho| < 0.4$, $|\rho| < 0.6$, and $|\rho| < 0.8$ (equivalent to $r^2 < 0.16$, $r^2 < 0.36$, and $r^2 < 0.64$). We note that pruning at $|\rho| < 0.1$ would often result in 3 or fewer variants being available for analysis, which would not allow multivariable Mendelian randomization to be attempted, as the number of genetic variants needs to be greater than the number of traits. Pruning is performed by first selecting the variant with the lowest p-value for association with any of the traits, and then excluding from consideration all variants more strongly correlated with the selected variant than the threshold value. We then select the variant amongst those remaining with the lowest p-value, and exclude variants more correlated with that variant than the threshold value. We repeat until all variants have either been selected or excluded. We also consider an oracle choice of variants, in which only the 15 genetic variants that truly influence the traits are used as instruments.

In addition to the main simulation study, we also consider the performance of methods with other parameter settings: 1) weaker instruments: we generate the $\alpha_{j}$ parameters from a normal distribution with mean 0.5 (corresponding to a mean instrument strength of $R^2 = 1.4\%$, mean univariable F = $57$, mean conditional F = $9.4$); 2) stronger correlations: we generate elements of the $A$ matrix from a uniform distribution on $+0.1$ to $+1.0$, resulting in correlations which typically range from around $+0.5$ to $+0.85$ with an interquartile range from around $+0.65$ to $+0.75$; 3) stronger causal effects, with $\theta_1 = +0.8$ and $\theta_3 = +1.0$, and 4) two alternative approaches for generating a correlation matrix taken from the R package \emph{clusterGeneration} \citep{joe2006}. For each scenario, 10\thinspace000 datasets were generated.

Further, we consider two variations to the main simulation study that are reflective of potential problems that may arise in applied practice. First, we estimate the variant correlation matrix based on an independent sample. This reflects the common occurrence that the correlation matrix is obtained from a reference sample rather than the dataset under analysis, and assesses robustness of the methods to variability in the correlation matrix. Secondly, we round the genetic associations and their standard errors to three decimal places. This reflects the common occurrence that genetic associations are obtained from a publicly-available source, and hence are not known to absolute precision. Again, this assesses robustness of the methods to variability in the data inputs.

\subsection*{Applied example: chemokine gene cluster and risk of stroke}
We illustrate our methods using data on genetic associations with three cytokines and stroke risk. Previous research has implicated monocyte chemoattractant protein-1 (MCP-1), which is also called chemokine (C-C motif) ligand 2 (CCL2), in the pathophysiology of stroke \citep{georgakis2019, georgakis2021basic, georgakis2021epi}. However, the \emph{CCL2} gene that encodes this cytokine is located in a cluster that also includes genes \emph{CCL7}, and \emph{CCL11}. Variants in this genetic region are associated with multiple cytokines other than MCP-1, including MCP-3 (also called CCL7), and eotaxin-1 (also called CCL11). Hence, it is not clear from univariable Mendelian randomization (that is, analyses with a single exposure trait) which of these proteins is driving stroke risk.

We conduct a multivariable cis-Mendelian randomization analysis to disentangle the effects of these cytokines. We take variants from the \emph{CCL2} gene region (GRCh38/hg38, chr17:34\thinspace255\thinspace218\thinspace-\thinspace34\thinspace257\thinspace203) plus 500 kilobasepairs either side, genetic associations with the cytokines from a re-analysis of data on three Finnish cohorts by Ahola-Olli \emph{et al} \citep{ahola2017} that did not adjust for body mass index \citep{kalaoja2021}, and genetic associations with all stroke and cardioembolic stroke from the MEGASTROKE consortium \citep{malik2018}. Cardioembolic stroke was chosen as genetic associations were stronger with this stroke subtype than for all stroke in a motivating Mendelian randomization analysis that included variants from throughout the genome \citep{georgakis2019}. Correlations between variants were estimated in 376\thinspace703 individuals of European ancestries from UK Biobank.  

\section*{Results}
\subsection*{Simulation study}
Results from the simulation study are shown in Table~\ref{tab:main}. For each method, we display the mean estimate of each parameter, the standard deviation of estimates, the mean standard error, and the empirical power of the 95\% confidence interval, which represents the proportion of confidence intervals that exclude the null. For $\theta_2$, the empirical power is the Type 1 error rate, and should be close to 5\%.

While the MV-IVW and MV-LIML methods perform well under the oracle setting, power to detect a causal effect is substantially reduced when pruning variants at 0.4. Although increasing the pruning threshold to 0.6 increases power, it also results in Type 1 error inflation. For the MV-IVW method, Type 1 error rates increase to 9.1\%, and for the MV-LIML method, to 20.2\%. When increasing the pruning threshold to 0.8, estimates are completely unreliable, with mean estimates in the MV-IVW method for $\theta_1$ and $\theta_3$ having the wrong sign.

In contrast, the MV-IVW-PCA and MV-LIML-PCA methods perform well throughout, with Type 1 error rates similar to those of the oracle methods, and greater power to detect a causal effect than the methods that rely on pruning. For the MV-IVW-PCA method, the variability and mean standard errors of estimates are similar to those of the oracle methods, although estimates of $\theta_1$ and $\theta_3$ are attenuated. This is a result of weak instrument bias \citep{burgess2016overlap}. For the MV-LIML-PCA method, estimates are less attenuated, but slightly more variable with greater mean standard errors. Compared with the MV-IVW-PCA method, power to detect a causal effect is slightly increased, although the Type 1 error rate was slightly higher (8.7\% versus 7.1\%).

Similar findings were observed when considering weaker instruments (Supplementary Table~\ref{tab:weak}), stronger correlations (Supplementary Table~\ref{tab:strong}), stronger causal effects (Supplementary Table~\ref{tab:strongeff}), and alternative correlation matrices (Supplementary Tables~\ref{tab:vine} and \ref{tab:onion}). Although the power varied between simulation settings, in each case the PCA methods outperformed the pruning methods in terms of power and precision at a threshold of 0.4, whereas at a threshold of 0.6 the MV-IVW and MV-LIML methods had inflated Type 1 error rates. Type 1 error rates for the PCA methods were generally well controlled, although the Type 1 error for the MV-LIML-PCA method was slightly inflated with weaker instruments (6.7\% for MV-IVW-PCA, 13.9\% for MV-LIML-PCA), and the Type 1 error for the MV-IVW-PCA method was slightly inflated with stronger causal effects (12.8\% for MV-IVW-PCA, 8.2\% for MV-LIML-PCA).

We also considered two additional variations to the main simulation reflective of potential problems in applied practice. Table~\ref{tab:diffcorr} shows results in which the variant correlation matrix was obtained from an independent sample of 10\thinspace000 individuals. Results were similar, except that the Type 1 error rate for the MV-IVW method at a pruning threshold of 0.6 was slightly higher at 11.2\%. When obtaining the correlation matrix from an independent sample of 1000 individuals, Type 1 error rates for the MV-IVW method were higher still at 18.7\% (Supplementary Table~\ref{tab:diffcorr2}). Table~\ref{tab:round} shows results in which the genetic association estimates were rounded to 3 decimal places. Again, results were similar, except that the Type 1 error rate for the MV-IVW method at a pruning threshold of 0.6 was notably higher at 15.1\%. In contrast, results from the PCA methods were not sensitive to changes in the variant correlation matrix or rounding of the genetic association estimates.

\subsection*{Applied example: chemokine gene cluster and risk of stroke}
Genetic associations with each of the cytokines and all stroke were available for 2922 variants, and with cardioembolic stroke for 2904 variants. We compare results from the MV-IVW-PCA method to those from the MV-IVW method at a pruning threshold of $|\rho| < 0.1$ (equivalent to $r^2 < 0.01$), $|\rho| < 0.4$ (equivalent to $r^2 < 0.16$), and $|\rho| < 0.6$ (equivalent to $r^2 < 0.13$). In the MV-IVW-PCA method, we initially pruned at $|\rho| < 0.95$ to remove very highly correlated variants from the analysis. We also excluded variants not associated with any of the cytokines at $p<0.001$ from all analyses. Estimates for the three cytokines, which represent log odds ratios per 1 standard deviation increase in the cytokine, are provided in Table~\ref{tab:chemokine}.

For all stroke, at a pruning threshold of $0.1$, the MV-IVW method indicates that MCP-1 is the true causal risk factor. However, a researcher may be tempted to consider a less strict pruning threshold to obtain more precise estimates. But at a pruning threshold of $0.4$, the MV-IVW method indicates that eotaxin-1 is the true causal risk factor, and at a pruning threshold of $0.6$, the MV-IVW method again indicates that MCP-1 has the strongest evidence of being the true causal risk factor, but the causal estimate is in the other direction. In contrast, the MV-IVW-PCA method indicates that MCP-1 has the strongest evidence of being the true causal risk factor, similarly to the MV-IVW method at the most conservative pruning threshold. Compared with results at this threshold, estimates from the MV-IVW-PCA method have narrower standard errors, although the estimate for MCP-1 is slightly attenuated and so has a slightly higher p-value. At a pruning threshold of 0.4, the condition number for the variance-covariance matrix $\Sigma$ in the MV-IVW method is 1224, whereas the condition number for the transformed variance-covariance matrix $\tilde{\Sigma}$ in the MV-IVW-PCA method is 24.9; a larger number signals worse problems due to ill-conditioning. We would therefore trust results from the MV-IVW-PCA method and the MV-IVW method at a threshold of 0.1, which both suggest that the strongest evidence is for MCP-1 as the causal risk factor, and the effect is in the harmful direction. For cardioembolic stroke, estimates are more similar amongst the different implementations of the methods, with stronger evidence for MCP-1 as the causal risk factor at this locus, particularly from the MV-IVW-PCA method. These findings add to the existing body of basic science \citep{georgakis2021basic}, observational \citep{georgakis2021epi, georgakis2019epi}, and genetic evidence \citep{georgakis2019} implicating circulating MCP-1 in stroke risk.

\section*{Discussion}
In this manuscript, we have introduced two methods for multivariable cis-Mendelian randomization, the MV-IVW-PCA and MV-LIML-PCA methods. Compared to existing methods that rely on pruning, these methods had superior performance: they outperformed pruning methods in terms of power to detect a causal effect, and they generally maintained close to nominal Type 1 error rates across a range of scenarios. They were also less sensitive that the pruning methods to variation in the variant correlation matrix, and to rounding of the genetic association estimates. We applied the MV-IVW-PCA method to disentangle the effects of three similar exposures with shared genetic predictors at a gene cluster; the method gave results that correspond to existing biological understanding of this pathway.

The approach of multivariable cis-Mendelian randomization has several potential applications. In our applied example, we considered proteins as exposures. Alternative potential applications could include expression of different genes as exposures, or expression of the same gene in different tissues. However, results from the latter case may be difficult to interpret if the genetic predictors of gene expression do not vary between tissues, or if data on variants affecting gene expression in all relevant tissues are not available. Another possible area of application is if there are different aspects of an exposure trait that could be considered as independent risk factors, such as concentration and isoform size of lipoprotein(a) \citep{saleheen2017}. To obtain estimates for the different exposures, it is not necessary to have genetic predictors that are uniquely associated with each exposure trait, but it is necessary to have some variants that associate more or less strongly with the different traits \citep{sanderson2018}.

An alternative approach for disentangling clusters of correlated variants associated with multiple traits is colocalization. Colocalization is a method that attempts to distinguish between two scenarios: a scenario in which two traits (which we here conceptualize as an exposure and an outcome) are influenced by distinct genetic variants, but there is overlap in the genetic associations due to correlation between the variants; and an alternative scenario in which the two traits are influenced by the same genetic variant \citep{wallace2012}. In the latter scenario, it is likely that the two traits are causally related, although the colocalization method is agnostic as to whether the exposure trait influences the outcome trait, the outcome influences the exposure, or the two are influenced by a common causal factor \citep{solovieff2013}. There are several conceptual differences between colocalization and cis-Mendelian randomization, although there are also similarities. Two specific advantages of the proposed cis-Mendelian randomization method are that it allows for the existence of multiple causal variants, in contrast to some colocalization methods \citep{giambartolomei2014, foley2019}, and it allows for the existence of multiple causal traits. Another feature is that it provides causal estimates, although the value of causal estimates in Mendelian randomization beyond indicating the direction of effect is disputed \citep{vanderweele2014}.

Although the PCA methods can be implemented with highly correlated variants, we would recommend a minimal level of pruning (say, $r^2<0.9$) before applying the methods in practice as variants that are very highly correlated do not contribute independent information to the analysis, but could provide computational challenges. Additionally, we have assumed that the traits under analysis are causally independent. If a trait has a causal effect on the outcome that is fully mediated by one of the other traits in the analysis, then the estimate for that trait would be zero. Hence, the method identifies the proximal causal risk factors for the outcome \citep{grant2020mvmr}. 
If there are large numbers of traits, the MV-IVW-PCA method could be combined with a Bayesian variable selection method that compares models with different sets of traits on the assumption of a sparse risk factor set \citep{zuber2018}.

There are several limitations to these methods, which are shared by other methods for Mendelian randomization using summarized data \citep{bowden2017, burgess2015scoretj}. Uncertainty in genetic associations with the exposure traits is not accounted for in the analysis. However, this is typically small compared with uncertainty in the genetic associations with the outcome, as variants selected for inclusion in the analysis are typically associated with at least one of the traits at a robust level of statistical significance. The effects of the exposure traits on the outcome are assumed to be linear. This is usually a reasonable assumption, given the small influence of genetic variants on traits, meaning that estimates reflect average causal effects for a small shift in the overall distribution of a trait \citep{burgess2014nonlin}. In our main simulation, all the exposure and outcome traits are continuous. For binary traits, the method can be implemented using genetic association estimates obtained from logistic regression. We have previously shown that multivariable Mendelian randomization methods are still able to make correct inferences in this setting \citep{burgess2014pleioaje, grant2020mvmr}, although the interpretation of estimates is obscured due to non-collapsibility \citep{burgess2012noncollapse}. Finally, estimates are subject to weak instrument bias. Whereas in univariable Mendelian randomization, weak instrument bias in a two-sample setting is towards the null \citep{pierce2012}, in multivariable Mendelian randomization, weak instruments can bias estimates in any direction \citep{zuber2018}. This is because weak instrument bias is analogous to measurement error, as the genetically-predicted values of the exposures are estimated with uncertainty, which in a multivariable regression model can lead to arbitrary bias \citep{thouless1939, phillips1991}. Hence it is important to balance the inclusion of several genetic variants in the analysis to achieve identification, with the inclusion of only variants strongly associated with the exposures to minimize weak instruments. The optimal balance will depend on the specifics of the analysis (such as the sample size available), but researchers should consider the conditional strength of instruments (via conditional F statistics) as well as the more conventional univariable F statistics \citep{sanderson2018}. Performance with weak instruments was mixed; mean estimates from the MV-LIML-PCA method were generally less affected than those from the MV-IVW-PCA method, although both methods had slightly elevated Type 1 error rates in one of the scenarios considered. We therefore recommend that both methods are applied when the instruments are weak, and caution is expressed if the methods give divergent results. Overall, we slightly prefer the MV-IVW-PCA method, as the Type 1 error rates from this method were generally slightly lower.

In summary, multivariable cis-Mendelian randomization can be used to disentangle the causal relationships of traits, such as proteins or gene expression measurements, that are influenced by a cluster of correlated genetic variants. The proposed PCA methods provide a compromise between loss of precision resulting from over-pruning and numerical instability resulting from under-pruning, to allow valid statistical tests that identify the causal traits influencing the outcome.

\section*{Data availability statement}
The summary statistics for genetic associations with three cytokines that support the findings of this study are available through \citep{ahola2017}. The summary statistics for genetic associations with any stroke and cardioembolic stroke are available in MEGASTROKE consortium \citep{malik2018} and data were derived from the public domain at [\url{www.megastroke.org}].  The variants correlation were accessed from UK Biobank \citep{biobank2014uk} at [\url{www.ukbiobank.ac.uk}].

\bibliographystyle{apalike}
\bibliography{cismvmrfb.bib}

\captionsetup[table]{skip=-10pt}
\begin{table}[hbtp]
\caption{Results from the main simulation study} 
\begin{center}
\begin{threeparttable}
\begin{tabular}[c]{cccccccccc}
\hline
Parameter  & Method  & Pruning &   Mean   &  SD   & Mean SE & Power  \\
\hline
$\theta_1$ & MV-IVW  & Oracle  &  $0.353$ & 0.133 & 0.120   & 81.4   \\
           &         & 0.4     &  $0.304$ & 0.164 & 0.147   & 57.4   \\
           &         & 0.6     &  $0.207$ & 0.115 & 0.094   & 60.9   \\
           &         & 0.8     & $-0.083$ & 0.417 & 0.051   & 76.5   \\
           & MV-LIML & Oracle  &  $0.379$ & 0.143 & 0.133   & 81.4   \\
           &         & 0.4     &  $0.340$ & 0.188 & 0.163   & 58.9   \\
           &         & 0.6     &  $0.316$ & 0.212 & 0.103   & 77.0   \\
           &         & 0.8     &  $0.083$ & 2.372 & 0.179   & 78.8   \\
           & MV-IVW-PCA  & -   &  $0.296$ & 0.130 & 0.119   & 69.3   \\
           & MV-LIML-PCA & -   &  $0.347$ & 0.152 & 0.130   & 74.5   \\
\hline
$\theta_2$ & MV-IVW  & Oracle  & $-0.005$ & 0.133 & 0.120   &  7.4   \\
           &         & 0.4     & $-0.012$ & 0.166 & 0.147   &  7.5   \\
           &         & 0.6     & $-0.003$ & 0.112 & 0.094   &  9.1   \\
           &         & 0.8     &  $0.037$ & 0.408 & 0.051   & 75.4   \\
           & MV-LIML & Oracle  & $-0.001$ & 0.144 & 0.133   &  6.2   \\
           &         & 0.4     & $-0.007$ & 0.192 & 0.163   &  7.3   \\
           &         & 0.6     &  $0.006$ & 0.186 & 0.103   & 20.2   \\
           &         & 0.8     &  $0.010$ & 2.522 & 0.179   & 76.6   \\
           & MV-IVW-PCA  & -   & $-0.013$ & 0.129 & 0.119   &  7.1   \\
           & MV-LIML-PCA & -   & $-0.006$ & 0.154 & 0.130   &  8.7   \\
\hline
$\theta_3$ & MV-IVW  & Oracle  & $-0.545$ & 0.132 & 0.120   & 98.6   \\
           &         & 0.4     & $-0.487$ & 0.166 & 0.148   & 87.6   \\
           &         & 0.6     & $-0.315$ & 0.139 & 0.094   & 86.9   \\
           &         & 0.8     &  $0.220$ & 0.418 & 0.051   & 77.8   \\
           & MV-LIML & Oracle  & $-0.576$ & 0.142 & 0.134   & 98.8   \\
           &         & 0.4     & $-0.531$ & 0.190 & 0.164   & 88.6   \\
           &         & 0.6     & $-0.451$ & 0.212 & 0.103   & 92.6   \\
           &         & 0.8     &  $0.005$ & 2.250 & 0.179   & 79.3   \\
           & MV-IVW-PCA  & -   & $-0.476$ & 0.130 & 0.119   & 96.1   \\
           & MV-LIML-PCA & -   & $-0.538$ & 0.152 & 0.131   & 97.0    \\
\hline
\end{tabular} 
\begin{tablenotes}
 Mean estimates, standard deviation (SD) of estimates, mean standard error (mean SE) of estimates, and empirical power of the 95\% confidence interval to estimate $\theta_1 = 0.4$, $\theta_2 = 0$, and $\theta_3 = -0.6$. We consider four methods, and various pruning thresholds for the MV-IVW and MV-LIML methods, plus an oracle setting in which only the 15 variants that truly affect the traits are included in the analysis.
\end{tablenotes}
\end{threeparttable}
\end{center}
\label{tab:main}
\end{table}

\captionsetup[table]{skip=-10pt}
\begin{table}[hbtp]
\caption{Results from the main simulation study with a correlation matrix estimated in an independent sample of 10\thinspace000 individuals} 
\begin{center}
\begin{threeparttable}
\begin{tabular}[c]{cccccccccc}
\hline
Parameter  & Method  & Pruning &   Mean   &  SD   & Mean SE & Power  \\
\hline
$\theta_1$ & MV-IVW  & 0.4     &  $0.303$ & 0.166 & 0.147   & 57.1   \\
           &         & 0.6     &  $0.206$ & 0.114 & 0.090   & 62.4   \\
           & MV-LIML & 0.4     &  $0.340$ & 0.189 & 0.162   & 59.0   \\
           &         & 0.6     &  $0.301$ & 0.183 & 0.098   & 76.7   \\
           & MV-IVW-PCA  & -   &  $0.295$ & 0.130 & 0.118   & 69.2   \\
           & MV-LIML-PCA & -   &  $0.347$ & 0.153 & 0.130   & 74.7   \\
\hline
$\theta_2$ & MV-IVW  & 0.4     & $-0.013$ & 0.167 & 0.148   &  7.3   \\
           &         & 0.6     & $-0.005$ & 0.114 & 0.090   & 11.2   \\
           & MV-LIML & 0.4     & $-0.009$ & 0.193 & 0.163   &  7.3   \\
           &         & 0.6     &  $0.003$ & 0.174 & 0.098   & 20.8   \\
           & MV-IVW-PCA  & -   & $-0.012$ & 0.129 & 0.118   &  7.2   \\
           & MV-LIML-PCA & -   & $-0.006$ & 0.154 & 0.130   &  8.9   \\
\hline
$\theta_3$ & MV-IVW  & 0.4     & $-0.485$ & 0.165 & 0.148   & 86.8   \\
           &         & 0.6     & $-0.322$ & 0.126 & 0.090   & 88.9   \\
           & MV-LIML & 0.4     & $-0.528$ & 0.188 & 0.164   & 87.9   \\
           &         & 0.6     & $-0.436$ & 0.207 & 0.099   & 93.6   \\
           & MV-IVW-PCA  & -   & $-0.476$ & 0.130 & 0.119   & 96.1   \\
           & MV-LIML-PCA & -   & $-0.538$ & 0.152 & 0.131   & 97.0   \\
\hline
\end{tabular}
\begin{tablenotes}
Mean estimates, standard deviation (SD) of estimates, mean standard error (mean SE) of estimates, and empirical power of the 95\% confidence interval to estimate $\theta_1 = 0.4$, $\theta_2 = 0$, and $\theta_3 = -0.6$. 
\end{tablenotes}
\end{threeparttable}
\label{tab:diffcorr}
\end{center}
\end{table}

\captionsetup[table]{skip=-10pt}
\begin{table}[hbtp]
\caption{Results from the main simulation study with genetic associations (beta-coefficients and standard errors) rounded to 3 decimal places}
\begin{center}
\begin{threeparttable}
\begin{tabular}[c]{cccccccccc}
\hline
Parameter  & Method  & Pruning &   Mean   &  SD   & Mean SE & Power  \\
\hline
$\theta_1$ & MV-IVW  & 0.4     &  $0.305$ & 0.165 & 0.147   & 57.6   \\
           &         & 0.6     &  $0.200$ & 0.138 & 0.090   & 59.7   \\
           & MV-LIML & 0.4     &  $0.341$ & 0.187 & 0.162   & 59.4   \\
           &         & 0.6     &  $0.302$ & 0.192 & 0.099   & 75.6   \\
           & MV-IVW-PCA  & -   &  $0.296$ & 0.130 & 0.119   & 69.2   \\
           & MV-LIML-PCA & -   &  $0.346$ & 0.153 & 0.130   & 74.7   \\
\hline
$\theta_2$ & MV-IVW  & 0.4     & $-0.013$ & 0.165 & 0.148   &  7.3   \\
           &         & 0.6     & $-0.012$ & 0.131 & 0.090   & 15.1   \\
           & MV-LIML & 0.4     & $-0.009$ & 0.190 & 0.163   &  7.1   \\
           &         & 0.6     & $-0.001$ & 0.183 & 0.099   & 24.4   \\
           & MV-IVW-PCA  & -   & $-0.013$ & 0.129 & 0.119   &  7.1   \\
           & MV-LIML-PCA & -   & $-0.006$ & 0.154 & 0.130   &  8.8   \\
\hline
$\theta_3$ & MV-IVW  & 0.4     & $-0.487$ & 0.166 & 0.148   & 87.5   \\
           &         & 0.6     & $-0.330$ & 0.145 & 0.090   & 89.3   \\
           & MV-LIML & 0.4     & $-0.529$ & 0.189 & 0.164   & 88.4   \\
           &         & 0.6     & $-0.459$ & 0.195 & 0.099   & 94.3   \\
           & MV-IVW-PCA  & -   & $-0.476$ & 0.131 & 0.119   & 96.0   \\
           & MV-LIML-PCA & -   & $-0.536$ & 0.152 & 0.131   & 96.9   \\
\hline
\end{tabular}
\begin{tablenotes}
Mean estimates, standard deviation (SD) of estimates, mean standard error (mean SE) of estimates, and empirical power of the 95\% confidence interval to estimate $\theta_1 = 0.4$, $\theta_2 = 0$, and $\theta_3 = -0.6$.
\end{tablenotes}
\end{threeparttable}
\label{tab:round}
\end{center}
\end{table}

 \captionsetup[table]{skip=-10pt}
\begin{table}[hbtp]
\caption{Applied example: effect of three cytokines on stroke risk }
\begin{center}
\begin{threeparttable}
\resizebox{\textwidth}{!}{%
{\small{
\begin{tabular}[c]{cccc|cc|cc|cc}
\hline
        &         & Variants /  & Cond   & \multicolumn{2}{c|}{MCP-1} & \multicolumn{2}{c|}{MCP-3} & \multicolumn{2}{c}{Eotaxin-1} \\
Method  & Pruning & PCs         & number & Estimate (SE)  & p-value   & Estimate (SE)    & p-value & Estimate (SE) & p-value    \\
\hline
\multicolumn{10}{c}{All stroke} \\
\hline
MV-IVW  & 0.1     &  20         & 27.7   & 0.091  (0.045) & 0.046     & $-0.062$ (0.046) & 0.18    & 0.062 (0.082) & 0.45       \\
        & 0.4     &  75         & 1224   & 0.057  (0.035) & 0.11      & $-0.014$ (0.024) & 0.55    & 0.110 (0.050) & 0.028      \\
        & 0.6     & 151         & 17762 & $-0.038$ (0.022) & 0.09     & $-0.014$ (0.017) & 0.41    & 0.040 (0.029) & 0.17       \\
\multicolumn{2}{l}{MV-IVW-PCA}  
                  &  30         & 24.9   & 0.075  (0.041) & 0.071     & $-0.029$ (0.027) & 0.29    & 0.000 (0.063) & 0.99       \\
\hline
\multicolumn{10}{c}{Cardioembolic stroke} \\
\hline
MV-IVW  & 0.1     &  19         & 22.7   & 0.270  (0.095) & 0.005     & 0.151  (0.104) & 0.14      & $-0.174$ (0.169) & 0.30    \\
        & 0.4     &  70         & 870    & 0.141  (0.073) & 0.053     & 0.003  (0.051) & 0.96      & $-0.132$ (0.107) & 0.21    \\
        & 0.6     & 145         & 15790  & 0.089  (0.046) & 0.056     & 0.040  (0.036) & 0.27      &   0.019 (0.062)  & 0.76    \\
\multicolumn{2}{l}{MV-IVW-PCA}  
                  &  29         & 25.9   & 0.254  (0.089) & 0.004     & $-0.018$ (0.065) & 0.78    & $-0.108$ (0.145) & 0.46    \\
\hline
\end{tabular}
}}}
\begin{tablenotes}
\setlength{\hsize}{0.80\hsize}
\justifying
\small
\noindent Estimates (standard errors, SE) and p-values from MV-IVW and MV-IVW-PCA methods. Variants/ PCs indicates the number of genetic variants (MV-IVW method) or principal components (PCs, MV-IVW-PCA method) included in the analysis. Cond number indicates the condition number of the variance-covariance matrix $\Sigma$; larger numbers signal worse problems due to ill-conditioning. Estimates represent log odds ratios  per 1 standard deviation increase in the cytokine.
\end{tablenotes}
\end{threeparttable}
\label{tab:chemokine}
\end{center}
\end{table}

\clearpage
\renewcommand{\thesection}{A\arabic{section}}
\renewcommand{\thesubsection}{A.\arabic{subsection}}
\renewcommand{\thetable}{A\arabic{table}}
\renewcommand{\thefigure}{A\arabic{figure}}
\renewcommand{\theequation}{A\arabic{equation}}
\setcounter{table}{0}
\setcounter{figure}{0}
\setcounter{equation}{0}
\renewcommand{\tablename}{Supplementary Table}
\renewcommand{\figurename}{Supplementary Figure}
\setcounter{section}{0}
\setcounter{subsection}{0}
\section*{Supplementary Material}
\subsection{Software code for MV-IVW and MV-IVW-PCA methods}
\normalsize{We provide R code to implement the MV-IVW and MV-IVW-PCA methods with three exposure traits. Genetic associations with the exposure are denoted \texttt{beta\_x1}, \texttt{beta\_x2}, and \texttt{beta\_x3}, and their standard errors \texttt{se\_x1}, \texttt{se\_x2}, and \texttt{se\_x3}. Genetic associations with the outcome are denoted \texttt{beta\_y} and their standard errors \texttt{se\_y}. The variant correlation matrix is denoted \texttt{rho}. Code is supplied both using the \emph{MendelianRandomization} package, and using matrix calculations directly:}
\footnotesize{
\begin{verbatim}
# MV-IVW method
###############
# note use of fixed-effect model under assumption that all variants in same
# genetic region and hence influence the outcome through same causal pathway

library(MendelianRandomization)
mvivw = mr_mvivw(mr_mvinput(cbind(beta_x1, beta_x2, beta_x3),
  cbind(se_x1, se_x2, se_x3), beta_y, se_y, correl=rho), model="fixed")
mvivw$Estimate; mvivw$StdError

beta_x    = cbind(beta_x1, beta_x2, beta_x3)
Sigma     = se_y%o%se_y*rho
mvivw_est = solve(t(beta_x)%*%solve(Sigma)%*%beta_x)%*%t(beta_x)%*%solve(Sigma)%*%beta_y
mvivw_se  = sqrt(diag(solve(t(beta_x)%*%solve(Sigma)%*%beta_x)))

# MV-PCA method
###############

Psi = ((abs(beta_x1)+abs(beta_x2)+abs(beta_x3))/se_y)%o%
      ((abs(beta_x1)+abs(beta_x2)+abs(beta_x3))/se_y)*rho
K   = which(cumsum(prcomp(Psi, scale=FALSE)$sdev^2/
               sum((prcomp(Psi, scale=FALSE)$sdev^2)))>0.99)[1]
       # K is number of principal components to include in analysis
       # this code includes principal components to explain 99% of variance in the risk factor
betaXG1   = as.numeric(beta_x1%*%prcomp(Psi, scale=FALSE)$rotation[,1:K])
betaXG2   = as.numeric(beta_x2%*%prcomp(Psi, scale=FALSE)$rotation[,1:K])
betaXG3   = as.numeric(beta_x3%*%prcomp(Psi, scale=FALSE)$rotation[,1:K])
betaYG0   = as.numeric(beta_y%*%prcomp(Psi, scale=FALSE)$rotation[,1:K])
sebetaYG0 = as.numeric(se_y%*%prcomp(Psi, scale=FALSE)$rotation[,1:K])
Sigma     = se_y%o%se_y*rho
pcSigma   = t(prcomp(Psi, scale=FALSE)$rotation[,1:K])%*%Sigma%*%
              prcomp(Psi, scale=FALSE)$rotation[,1:K]

mvpca     = mr_mvivw(mr_mvinput(cbind(betaXG1, betaXG2, betaXG3),
  cbind(rep(1, length(betaXG1)), rep(1, length(betaXG1)), rep(1, length(betaXG1))),
                     betaYG0, rep(1, length(betaXG1)), corr=pcSigma), model="fixed")
mvpca$Estimate; mvpca$StdError
# note that the standard errors of the genetic associations with the exposures are
# not used in the calculation, and so are set to 1

mvpca_est = solve(rbind(betaXG1, betaXG2, betaXG3)%*%solve(pcSigma)%*%
                  cbind(betaXG1, betaXG2, betaXG3))%*%
                  rbind(betaXG1, betaXG2, betaXG3)%*%solve(pcSigma)%*%betaYG0
mvpca_se  = sqrt(diag(solve(rbind(betaXG1, betaXG2, betaXG3)%*%solve(pcSigma)%*%
                            cbind(betaXG1, betaXG2, betaXG3))))
\end{verbatim}
}
\normalsize{\hphantom{end}}
\clearpage

\subsection{Software code for MV-LIML and MV-LIML-PCA methods}
\normalsize{We provide R code to implement the MV-LIML and MV-LIML-PCA methods. The matrix of genetic associations with the exposure is denoted \texttt{Bx}, and standard errors \texttt{Sx}. The vector of genetic associations with the outcome is denoted \texttt{By} and standard errors \texttt{Sy}. The variant correlation matrix is denoted \texttt{rho}, the trait correlation matrix is denoted \texttt{Phi}, and the number of principal components $R$; this can be provided by the user, or estimated automatically:}
\footnotesize{
\begin{verbatim}
# MV-LIML method
################

mvmr_liml <- function(Bx,By,Sx,Sy,rho,Phi) {
Bx <- as.matrix(Bx);   By <- as.vector(By)
Sx <- as.matrix(Sx);   Sy <- as.vector(Sy)
rho <- as.matrix(rho); Phi <- as.matrix(Phi)
J <- ncol(Bx); K <- nrow(Bx)

GamY <- (Sy%*%t(Sy))*rho

gamX <- list()
for (k in 1:J){
  gamX[[k]] <- (Sx[,k]%*%t(Sx[,k]))
}

GamX1 <- function(tet){
gamX.t  <- list()
for (k in 1:J){
  for (l in 1:J){
    gamX.t[[(((k-1)*J)+l)]] <- sqrt(gamX[[k]])*sqrt(gamX[[l]])*rho*tet[k]*tet[l]*Phi[k,l]
  }
}
return(gamX.t)
}

GamX <- function(tet){Reduce('+',GamX1(tet))}

# LIML estimation
g <- function(tet){as.vector(By - (Bx%*%tet))}
Om <- function(tet){as.matrix(GamY + GamX(tet))}
Q <- function(tet){as.numeric(t(g(tet))%*%(solve(Om(tet)))%*%g(tet))}
G <- -as.matrix(Bx)
DQ <- function(tet){2*as.matrix(t(G)%*%(solve(Om(tet)))%*%g(tet))}
liml <- nlminb(rep(0,J),objective=Q,gradient=DQ)
est <- as.vector(liml$par)
var.est <- as.matrix(solve(t(G)%*%(solve(Om(est)))%*%G))

res.list <- list("est"=est, "var"=var.est)
return(res.list)
}
\end{verbatim}
\clearpage
\begin{verbatim}
# MV-PCA-LIML method
####################

mv_pca_liml <- function(Bx,By,Sx,Sy,rho,Phi,R=NULL) {
Bx  <- as.matrix(Bx);   By <- as.vector(By)
Sx  <- as.matrix(Sx);   Sy <- as.vector(Sy)
rho <- as.matrix(rho); Phi <- as.matrix(Phi)
J <- ncol(Bx); K <- nrow(Bx)
Psi = (rowSums(abs(Bx))/Sy)%o%(rowSums(abs(Bx))/Sy)*rho

# estimate number of principal components
if(missing(R)) {
  R=which(cumsum(prcomp(Psi,scale=FALSE)$sdev^2/sum((prcomp(Psi,scale=FALSE)$sdev^2)))>0.99)[1]
} else {  R=R  }
lambda <- sqrt(K)*(eigen(Psi)$vectors[,1:R])
evec <- eigen((t(lambda)%*%lambda))$vectors
eval <- eigen((t(lambda)%*%lambda))$values
lambda <- lambda%*%(solve(evec%*%diag(sqrt(eval))%*%t(evec)))
dim(lambda) <- c(K,R)

GamY <- (Sy%*%t(Sy))*rho

gamX <- list()
for (k in 1:J){
  gamX[[k]] <- (Sx[,k]%*%t(Sx[,k]))
}

GamX1 <- function(tet){
gamX.t  <- list()
for (k in 1:J){
  for (l in 1:J){
    gamX.t[[(((k-1)*J)+l)]] <- sqrt(gamX[[k]])*sqrt(gamX[[l]])*rho*tet[k]*tet[l]*Phi[k,l]
  }
}
return(gamX.t)
}

GamX <- function(tet){Reduce('+',GamX1(tet))}

# LIML estimation
g <- function(tet){as.vector(t(lambda)%*%(By - (Bx%*%tet)))}
Om <- function(tet){as.matrix(t(lambda)%*%(GamY + GamX(tet))%*%lambda)}
Q <- function(tet){as.numeric(t(g(tet))%*%(solve(Om(tet)))%*%g(tet))}
G <- -as.matrix(t(lambda)%*%Bx)
DQ <- function(tet){2*as.matrix(t(G)%*%(solve(Om(tet)))%*%g(tet))}
liml <- nlminb(rep(0,J),objective=Q,gradient=DQ)
est <- as.vector(liml$par)
var.est <- as.matrix(solve(t(G)%*%(solve(Om(est)))%*%G))

res.list <- list("est"=est, "var"=var.est)
return(res.list)
}

\end{verbatim}
}
\normalsize{\hphantom{end}}
\clearpage

\subsection{Additional simulation results}
In addition to the main simulation study, we also consider the performance of the MV-IVW-PCA and MV-LIML-PCA methods with other parameter settings:
\begin{itemize}
  \item Supplementary Table~\ref{tab:weak}. Weaker instruments: we generate the $\alpha_{j}$ parameters from a normal distribution with mean 0.5.
  \item Supplementary Table~\ref{tab:strong}. Stronger correlations: we generate elements of the $A$ matrix from a uniform distribution on $+0.1$ to $+1.0$. Note that as the vast majority of correlations between variants were above +0.6, we did not attempt the methods that rely on pruning in this scenario.
    \item Supplementary Table~\ref{tab:strongeff}. Stronger effects: we set $\theta_1 = +0.8$ and $\theta_3 = +1.0$.
  \item Supplementary Table~\ref{tab:vine}. Alternative approach for generating a correlation matrix 1: the ``c-vine'' method from the R package \emph{clusterGeneration}. The code is: \newline \verb| sig <- cor(genPositiveDefMat(dim=vars, covMethod = "c-vine",| \newline \verb|    eigenvalue=runif(vars, -1, 1))$Sigma)|. \newline Correlations typically ranged from around $-0.5$ to $+0.5$ with an interquartile range from around $-0.1$ to $+0.1$.
  \item Supplementary Table~\ref{tab:onion}. Alternative approach for generating a correlation matrix 2: the ``onion'' method from the R package \emph{clusterGeneration}. The code is: \newline \verb| sig <- cor(genPositiveDefMat(dim=vars, covMethod = "onion",| \newline \verb|    eigenvalue=runif(vars, -1, 1))$Sigma)| \newline Correlations typically ranged from around $-0.5$ to $+0.5$ with an interquartile range from around $-0.1$ to $+0.1$.
  \item Supplementary Table~\ref{tab:diffcorr2}. Smaller independent sample for variant correlation matrix: the correlation matrix between genetic variants is estimates based on an independent sample of 1000 individuals.
\end{itemize}

In each setting, the proposed PCA methods performed well, with close to nominal Type 1 error rates in most scenarios and greater power than for the MV-IVW and MV-LIML methods at pruning thresholds where those methods maintained reasonable Type 1 error rates.

\captionsetup[table]{skip=-10pt}
\begin{table}[hbtp]
\caption{Results from simulation study with weaker instruments }
\begin{center}
\begin{threeparttable}
\begin{tabular}[c]{cccccccccc}
\hline
Parameter  & Method  & Pruning &   Mean   &  SD   & Mean SE & Power  \\
\hline
$\theta_1$ & MV-IVW  & Oracle  &  $0.297$ & 0.196 & 0.180   & 39.2   \\
           &         & 0.4     &  $0.215$ & 0.225 & 0.204   & 22.3   \\
           &         & 0.6     &  $0.119$ & 0.132 & 0.119   & 20.5   \\
           &         & 0.8     & $-0.050$ & 0.505 & 0.066   & 71.9   \\
           & MV-LIML & Oracle  &  $0.349$ & 0.237 & 0.201   & 43.0   \\
           &         & 0.4     &  $0.276$ & 0.312 & 0.229   & 28.8   \\
           &         & 0.6     &  $0.215$ & 0.266 & 0.128   & 42.5   \\
           &         & 0.8     &  $0.106$ & 2.425 & 0.234   & 75.2   \\
           & MV-IVW-PCA  & -   &  $0.208$ & 0.176 & 0.164   & 26.5   \\
           & MV-LIML-PCA & -   &  $0.296$ & 0.262 & 0.182   & 41.2   \\
\hline
$\theta_2$ & MV-IVW  & Oracle  & $-0.012$ & 0.197 & 0.180   &  6.9   \\
           &         & 0.4     & $-0.022$ & 0.228 & 0.205   &  7.0   \\
           &         & 0.6     & $-0.012$ & 0.134 & 0.119   &  8.0   \\
           &         & 0.8     & $-0.043$ & 0.499 & 0.065   & 72.0   \\
           & MV-LIML & Oracle  & $-0.006$ & 0.238 & 0.201   &  7.5   \\
           &         & 0.4     & $-0.015$ & 0.315 & 0.230   &  9.9   \\
           &         & 0.6     & $-0.002$ & 0.274 & 0.128   & 22.4   \\
           &         & 0.8     &  $0.015$ & 2.437 & 0.234   & 75.4   \\
           & MV-IVW-PCA  & -   & $-0.023$ & 0.176 & 0.164   &  6.7   \\
           & MV-LIML-PCA & -   & $-0.014$ & 0.264 & 0.182   & 13.9   \\
\hline
$\theta_3$ & MV-IVW  & Oracle  & $-0.478$ & 0.196 & 0.180   & 75.4   \\
           &         & 0.4     & $-0.381$ & 0.233 & 0.206   & 50.3   \\
           &         & 0.6     & $-0.204$ & 0.148 & 0.119   & 44.3   \\
           &         & 0.8     & $-0.181$ & 0.500 & 0.066   & 72.8   \\
           & MV-LIML & Oracle  & $-0.539$ & 0.233 & 0.202   & 78.2   \\
           &         & 0.4     & $-0.455$ & 0.317 & 0.231   & 56.5   \\
           &         & 0.6     & $-0.326$ & 0.275 & 0.128   & 64.4   \\
           &         & 0.8     & $-0.012$ & 2.283 & 0.233   & 74.7   \\
           & MV-IVW-PCA  & -   & $-0.368$ & 0.179 & 0.165   & 61.3   \\
           & MV-LIML-PCA & -   & $-0.475$ & 0.263 & 0.182   & 72.3   \\
\hline
\end{tabular}
\begin{tablenotes}
Mean estimates, standard deviation (SD) of estimates, mean standard error (mean SE) of estimates, and  empirical power of the 95\% confidence interval to estimate $\theta_1 = 0.4$, $\theta_2 = 0$, and $\theta_3 = -0.6$. We consider four methods, and various pruning thresholds for the MV-IVW and MV-LIML methods, plus an oracle setting in which only the 15 variants that truly affect the traits are included in the analysis.
 \end{tablenotes}
\end{threeparttable}
\label{tab:weak}
\end{center}
\end{table}

\begin{table}[hbtp]
\caption{Results from simulation study with stronger correlations}
\begin{center}
\begin{threeparttable}
\begin{tabular}[c]{cccccccccc}
\hline
Parameter  & Method  & Pruning &   Mean   &  SD   & Mean SE & Power  \\
\hline
$\theta_1$ & MV-IVW  & Oracle  &  $0.308$ & 0.182 & 0.166   & 47.9   \\
           & MV-LIML & Oracle  &  $0.345$ & 0.206 & 0.184   & 48.9   \\
           & MV-IVW-PCA  & -   &  $0.217$ & 0.166 & 0.153   & 32.2   \\
           & MV-LIML-PCA & -   &  $0.273$ & 0.219 & 0.167   & 39.6   \\
\hline
$\theta_2$ & MV-IVW  & Oracle  &  $-0.014$ & 0.182 & 0.166  &  7.2   \\
           & MV-LIML & Oracle  &  $-0.009$ & 0.209 & 0.185  &  6.8   \\
           & MV-IVW-PCA  & -   &  $-0.027$ & 0.165 & 0.154  &  7.1   \\
           & MV-LIML-PCA & -   &  $-0.020$ & 0.222 & 0.167  & 11.1   \\
\hline
$\theta_3$ & MV-IVW  & Oracle  &  $-0.493$ & 0.181 & 0.166  & 82.7   \\
           & MV-LIML & Oracle  &  $-0.535$ & 0.204 & 0.185  & 84.1   \\
           & MV-IVW-PCA  & -   &  $-0.387$ & 0.167 & 0.154  & 70.7   \\
           & MV-LIML-PCA & -   &  $-0.452$ & 0.223 & 0.168  & 75.6   \\
\hline
\end{tabular}
\begin{tablenotes}
Mean estimates, standard deviation (SD) of estimates, mean standard error (mean SE) of estimates, and empirical power of the 95\% confidence interval to estimate $\theta_1 = 0.4$, $\theta_2 = 0$, and $\theta_3 = -0.6$. Results for different pruning thresholds are omitted as the correlations between variants were typically all above 0.6.
\end{tablenotes}
\end{threeparttable}
 \label{tab:strong}
\end{center}
\end{table} 
 
\begin{table}[hbtp]
\caption{Results from simulation study with stronger effects}
\begin{center}
\begin{threeparttable}
\begin{tabular}[c]{cccccccccc}
\hline
Parameter  & Method  & Pruning &   Mean   &  SD   & Mean SE & Power  \\
\hline
$\theta_1$ & MV-IVW  & Oracle  &  $0.771$ & 0.214 & 0.185   & 96.7   \\
           &         & 0.4     &  $0.745$ & 0.265 & 0.226   & 87.2   \\
           &         & 0.6     &  $0.441$ & 0.341 & 0.144   & 81.1   \\
           &         & 0.8     & $-0.565$ & 0.994 & 0.078   & 88.5   \\
           & MV-LIML & Oracle  &  $0.786$ & 0.230 & 0.212   & 94.6   \\
           &         & 0.4     &  $0.767$ & 0.312 & 0.261   & 82.5   \\
           &         & 0.6     &  $0.665$ & 1.787 & 0.202   & 81.3   \\
           &         & 0.8     &  $0.407$ & 6.488 & 0.574   & 87.8   \\
           & MV-IVW-PCA  & -   &  $0.735$ & 0.209 & 0.182   & 95.9   \\
           & MV-LIML-PCA & -   &  $0.772$ & 0.257 & 0.209   & 92.0   \\
\hline
$\theta_2$ & MV-IVW  & Oracle  & $ 0.048$ & 0.209 & 0.185   &  9.0   \\
           &         & 0.4     & $ 0.108$ & 0.265 & 0.227   & 11.5   \\
           &         & 0.6     & $ 0.015$ & 0.295 & 0.144   & 28.0   \\
           &         & 0.8     & $-0.353$ & 1.011 & 0.078   & 86.7   \\
           & MV-LIML & Oracle  &  $0.020$ & 0.220 & 0.213   &  5.3   \\
           &         & 0.4     &  $0.061$ & 0.306 & 0.263   &  8.0   \\
           &         & 0.6     & $-0.355$ & 1.871 & 0.203   & 46.0   \\
           &         & 0.8     & $-1.306$ & 6.725 & 0.575   & 89.0   \\
           & MV-IVW-PCA  & -   &  $0.116$ & 0.205 & 0.182   & 12.8   \\
           & MV-LIML-PCA & -   &  $0.047$ & 0.236 & 0.209   &  8.2   \\
\hline
$\theta_3$ & MV-IVW  & Oracle  &  $0.956$ & 0.215 & 0.185   & 99.4   \\
           &         & 0.4     &  $0.904$ & 0.269 & 0.228   & 93.4   \\
           &         & 0.6     &  $0.542$ & 0.357 & 0.144   & 87.0   \\
           &         & 0.8     & $-0.618$ & 0.987 & 0.078   & 87.7   \\
           & MV-LIML & Oracle  &  $0.982$ & 0.229 & 0.213   & 99.0   \\
           &         & 0.4     &  $0.947$ & 0.319 & 0.263   & 91.4   \\
           &         & 0.6     &  $0.934$ & 1.912 & 0.202   & 89.7   \\
           &         & 0.8     &  $1.074$ & 6.675 & 0.575   & 88.5   \\
           & MV-IVW-PCA  & -   &  $0.891$ & 0.210 & 0.183   & 98.9   \\
           & MV-LIML-PCA & -   &  $0.954$ & 0.251 & 0.210   & 97.8   \\
\hline
\end{tabular}
\begin{tablenotes}
Mean estimates, standard deviation (SD) of estimates, mean standard error (mean SE) of estimates, and empirical power of the 95\% confidence interval to estimate $\theta_1 = 0.8$, $\theta_2 = 0$, and $\theta_3 = 1.0$. We consider four methods, and various pruning thresholds for the MV-IVW and MV-LIML methods, plus an oracle setting in which only the 15 variants that truly affect the traits are included in the analysis. 
\end{tablenotes}
\end{threeparttable}
\label{tab:strongeff} 
\end{center}
\end{table}

\begin{table}[hbtp]
\caption{Results from simulation study with alternative correlation matrix (``c-vine'')}
\begin{center}
\begin{threeparttable}
\begin{tabular}[c]{cccccccccc}
\hline
Parameter  & Method  & Pruning &   Mean   &  SD   & Mean SE & Power  \\
\hline
$\theta_1$ & MV-IVW  & Oracle  &  $0.371$ & 0.137 & 0.123   & 83.2   \\
           &         & 0.4     &  $0.274$ & 0.114 & 0.104   & 73.1   \\
           &         & 0.6     & $-0.135$ & 0.422 & 0.055   & 75.6   \\
           &         & 0.8     & $-0.143$ & 0.432 & 0.054   & 76.7   \\
           & MV-LIML & Oracle  &  $0.386$ & 0.144 & 0.137   & 80.9   \\
           &         & 0.4     &  $0.351$ & 0.152 & 0.115   & 80.9   \\
           &         & 0.6     &  $0.131$ & 1.019 & 0.114   & 72.8   \\
           &         & 0.8     &  $0.131$ & 1.035 & 0.115   & 73.5   \\
           & MV-IVW-PCA  & -   &  $0.334$ & 0.135 & 0.122   & 76.2   \\
           & MV-LIML-PCA & -   &  $0.368$ & 0.153 & 0.135   & 76.6   \\
\hline
$\theta_2$ & MV-IVW  & Oracle  &  $0.002$ & 0.135 & 0.123   &  7.3   \\
           &         & 0.4     &  $0.002$ & 0.113 & 0.104   &  6.9   \\
           &         & 0.6     &  $0.006$ & 0.414 & 0.055   & 74.4   \\
           &         & 0.8     &  $0.005$ & 0.423 & 0.054   & 75.5   \\
           & MV-LIML & Oracle  &  $0.001$ & 0.143 & 0.137   &  5.6   \\
           &         & 0.4     &  $0.001$ & 0.154 & 0.115   & 13.5   \\
           &         & 0.6     &  $0.002$ & 1.056 & 0.114   & 71.7   \\
           &         & 0.8     &  $0.007$ & 1.104 & 0.115   & 71.9   \\
           & MV-IVW-PCA  & -   &  $0.001$ & 0.134 & 0.122   &  6.8   \\
           & MV-LIML-PCA & -   &  $0.000$ & 0.153 & 0.135   &  7.6   \\
\hline
$\theta_3$ & MV-IVW  & Oracle  & $-0.558$ & 0.136 & 0.123   & 98.8   \\
           &         & 0.4     & $-0.411$ & 0.115 & 0.104   & 95.4   \\
           &         & 0.6     &  $0.212$ & 0.419 & 0.055   & 77.9   \\
           &         & 0.8     & $-0.226$ & 0.427 & 0.054   & 79.2   \\
           & MV-LIML & Oracle  & $-0.581$ & 0.141 & 0.136   & 98.7   \\
           &         & 0.4     & $-0.529$ & 0.146 & 0.115   & 97.6   \\
           &         & 0.6     & $-0.186$ & 0.934 & 0.113   & 75.2   \\
           &         & 0.8     & $-0.190$ & 0.970 & 0.115   & 75.5   \\
           & MV-IVW-PCA  & -   & $-0.502$ & 0.134 & 0.122   & 96.6   \\
           & MV-LIML-PCA & -   & $-0.555$ & 0.147 & 0.135   & 97.1   \\
\hline
\end{tabular}
\begin{tablenotes}
Mean estimates, standard deviation (SD) of estimates, mean standard error (mean SE) of estimates, and empirical power of the 95\% confidence interval to estimate $\theta_1 = 0.4$, $\theta_2 = 0$, and $\theta_3 = -0.6$. We consider four methods, and various pruning thresholds for the MV-IVW and MV-LIML methods, plus an oracle setting in which only the 15 variants that truly affect the traits are included in the analysis.
\end{tablenotes}
\end{threeparttable}
\label{tab:vine}
\end{center}
\end{table} 

\captionsetup[table]{skip=10pt}
\begin{table}[hbtp]
\begin{center}
\caption{Results from simulation study with alternative correlation matrix (``onion'')}
\begin{threeparttable}
\begin{tabular}[c]{cccccccccc}
\hline
Parameter  & Method  & Pruning &   Mean   &  SD   & Mean SE & Power  \\
\hline
$\theta_1$ & MV-IVW  & Oracle  &  $0.374$ & 0.136 & 0.123   & 83.6   \\
           &         & 0.4     &  $0.276$ & 0.114 & 0.105   & 73.6   \\
           &         & 0.6     & $-0.137$ & 0.421 & 0.055   & 75.8   \\
           &         & 0.8     & $-0.146$ & 0.434 & 0.054   & 77.2   \\
           & MV-LIML & Oracle  &  $0.390$ & 0.142 & 0.137   & 81.5   \\
           &         & 0.4     &  $0.354$ & 0.151 & 0.115   & 81.5   \\
           &         & 0.6     &  $0.127$ & 1.015 & 0.114   & 73.8   \\
           &         & 0.8     &  $0.127$ & 1.054 & 0.115   & 74.5   \\
           & MV-IVW-PCA  & -   &  $0.336$ & 0.134 & 0.122   & 76.5   \\
           & MV-LIML-PCA & -   &  $0.370$ & 0.151 & 0.135   & 77.5   \\
\hline
$\theta_2$ & MV-IVW  & Oracle  & $-0.000$ & 0.137 & 0.123   &  7.8   \\
           &         & 0.4     & $-0.000$ & 0.115 & 0.104   &  7.5   \\
           &         & 0.6     & $-0.000$ & 0.412 & 0.055   & 74.7   \\
           &         & 0.8     & $-0.002$ & 0.423 & 0.053   & 75.9   \\
           & MV-LIML & Oracle  & $-0.001$ & 0.145 & 0.137   &  6.0   \\
           &         & 0.4     & $-0.001$ & 0.159 & 0.115   & 14.6   \\
           &         & 0.6     &  $0.004$ & 1.007 & 0.113   & 72.0   \\
           &         & 0.8     &  $0.005$ & 1.036 & 0.115   & 72.8   \\
           & MV-IVW-PCA  & -   & $-0.000$ & 0.137 & 0.122   &  7.6   \\
           & MV-LIML-PCA & -   & $-0.001$ & 0.158 & 0.135   &  8.5   \\
\hline
$\theta_3$ & MV-IVW  & Oracle  & $-0.556$ & 0.134 & 0.123   & 98.7   \\
           &         & 0.4     & $-0.412$ & 0.114 & 0.104   & 95.5   \\
           &         & 0.6     &  $0.212$ & 0.431 & 0.055   & 78.0   \\
           &         & 0.8     &  $0.225$ & 0.441 & 0.053   & 79.3   \\
           & MV-LIML & Oracle  & $-0.584$ & 0.139 & 0.137   & 98.7   \\
           &         & 0.4     & $-0.531$ & 0.146 & 0.115   & 97.7   \\
           &         & 0.6     & $-0.189$ & 0.897 & 0.113   & 75.2   \\
           &         & 0.8     & $-0.186$ & 0.935 & 0.115   & 75.7   \\
           & MV-IVW-PCA  & -   & $-0.503$ & 0.133 & 0.122   & 96.6   \\
           & MV-LIML-PCA & -   & $-0.556$ & 0.146 & 0.135   & 97.2   \\
\hline
\end{tabular}
\begin{tablenotes}
Mean estimates, standard deviation (SD) of estimates, mean standard error (mean SE) of estimates, and empirical power of the 95\% confidence interval to estimate $\theta_1 = 0.4$, $\theta_2 = 0$, and $\theta_3 = -0.6$. We consider four methods, and various pruning thresholds for the MV-IVW and MV-LIML methods, plus an oracle setting in which only the 15 variants that truly affect the traits are included in the analysis.
\end{tablenotes}
\end{threeparttable}
\label{tab:onion}
\end{center}
\end{table}

\captionsetup[table]{skip=10pt}
\begin{table}[hbtp]
\begin{center}
\caption{Results from the main simulation study with a correlation matrix estimated in an independent sample of 1000 individuals}
\begin{threeparttable}
\begin{tabular}[c]{cccccccccc}
\hline
Parameter  & Method  & Pruning &   Mean   &  SD   & Mean SE & Power  \\
\hline
$\theta_1$ & MV-IVW  & 0.4     &  $0.304$ & 0.170 & 0.148   & 56.7   \\
           &         & 0.6     &  $0.213$ & 0.135 & 0.085   & 65.0   \\
           & MV-LIML & 0.4     &  $0.339$ & 0.199 & 0.165   & 58.0   \\
           &         & 0.6     &  $0.297$ & 0.191 & 0.093   & 77.1   \\
           & MV-IVW-PCA  & -   &  $0.297$ & 0.132 & 0.119   & 69.3   \\
           & MV-LIML-PCA & -   &  $0.347$ & 0.155 & 0.131   & 74.1   \\
\hline
$\theta_2$ & MV-IVW  & 0.4     & $-0.013$ & 0.170 & 0.149   &  8.0   \\
           &         & 0.6     & $-0.012$ & 0.136 & 0.085   & 18.7   \\
           & MV-LIML & 0.4     & $-0.009$ & 0.198 & 0.165   &  7.8   \\
           &         & 0.6     & $-0.002$ & 0.189 & 0.093   & 27.8   \\
           & MV-IVW-PCA  & -   & $-0.014$ & 0.134 & 0.119   &  7.8   \\
           & MV-LIML-PCA & -   & $-0.007$ & 0.160 & 0.131   &  9.6   \\
\hline
$\theta_3$ & MV-IVW  & 0.4     & $-0.487$ & 0.169 & 0.148   & 86.8   \\
           &         & 0.6     & $-0.355$ & 0.135 & 0.085   & 93.2   \\
           & MV-LIML & 0.4     & $-0.529$ & 0.193 & 0.164   & 87.5   \\
           &         & 0.6     & $-0.459$ & 0.189 & 0.093   & 95.7   \\
           & MV-IVW-PCA  & -   & $-0.478$ & 0.132 & 0.119   & 95.6   \\
           & MV-LIML-PCA & -   & $-0.538$ & 0.154 & 0.131   & 96.7   \\
\hline
\end{tabular}
\begin{tablenotes}
Mean estimates, standard deviation (SD) of estimates, mean standard error (mean SE) of estimates, and empirical power of the 95\% confidence interval to estimate $\theta_1 = 0.4$, $\theta_2 = 0$, and $\theta_3 = -0.6$.
\end{tablenotes}
\end{threeparttable}
\label{tab:diffcorr2}
\end{center}
\end{table}

\end{document}